This final part offers a survey of models proposed to cope with the symmetry-breaking challenge. Among them are the two-component neutrinos, the neutrino twins, the universal Fermi interaction, etc. Moreover, the broken discrete symmetries in physics are very much on the agenda and may occupy considerable time for LHC experiments aimed at revealing the symmetry-breaking mechanisms. Finally, an account of the achievements of dual-component theories in explaining parity-breaking phenomena is added.

### 7. Two-component neutrino

In the previous parts I through III of the paper we described the theoretical background as well as the bulk experimental evidence that discrete group symmetries break up in weak interactions, such as the β–decay. This last part IV will be devoted to the theoretical models designed to cope with the parity-breaking challenge. The quality of the proposed theories and the accuracy of the experiments made to check them underline the place occupied by papers such as ours in the dissemination of symmetry-related matter of modern science.

#### 7.1. Neutrino gauge

In the general form of β-interaction (5.6) the case

$$C_k' = \pm C_k \tag{7.1}$$

is of particular interest. It corresponds to interchanging the neutrino wave function in the parity-conserving Hamiltonian ( $C_k' = 0$ ) with the function

$$\psi_v' = (I \pm \gamma_s)\psi_v \tag{7.2}$$

In as much as parity is not conserved with this transformation, it is natural to put the blame on the neutrino. Note that the change (7.2) signifies that the neutrino wave function $\psi_v$ having a definite spatial parity $\eta_p$ another function is added to $\psi_v$ with the opposite parity $-\eta_p$. That is, the neutrino state (7.2) is a mixture of equal amounts of opposite parity states. The latter has the meaning of not attributing any definite spatial parity to the neutrino.

We will check just how the transformation (7.2) affects the properties of the free neutrino field. Dirac's equation of a free spinor particle has the (2.2) form. In as much as the function $\psi_v$ satisfies that equation, the invariance condition for the neutrino theory with respect to the change (7.2) is reduced to invariance with respect to the transformation

$$\psi_v \rightarrow \exp(i\alpha_v)\,\gamma_s\gamma_v \tag{7.3}$$

where $\exp(i\alpha_v)$ is a fixed phase factor specific for the internal structure of the neutrino field. This transform is called a neutrino gauge.

For the neutrino gauge the density of the Lagrangian of a free spinor field

$$L = -\underline{\psi}_v\,(\nabla^\wedge + m_v)\,\psi_v \tag{7.4}$$

goes to

$$L = -\underline{\psi}_v\,(\nabla^\wedge - m_v)\,\psi_v \tag{7.5}$$

and, consequently, the invariance condition is

$$m_v = 0 \tag{7.6}$$

that is, a vanishing inherent mass at rest of the neutrino field is required.

In this way the requirement for invariance of the neutrino theory with respect to the gauge transform (7.2) or all the same (7.3) leads to cancelling the rest mass of the free neutrino. This conclusion agrees with the experiment which indicates that in any event the free mass of the neutrino field is vanishingly small. We note that the above neutrino situation is similar to the one of the electromagnetic field where the gradient invariance of the field and all the interactions leads to the strict vanishing of the photon inherent rest mass.

In the search for a complete analogy, Salam has postulated that all interactions remain invariant on neutrino gauge. The general form of Fermi's weak interaction with neutrino participating (whereas the neutrino participates in weak interactions alone) is given by (5.2) provided we forget for a moment the specific meaning of the operators $P(x^\vee)$, $N(x^\vee)$, and $e(x^\vee)$. On neutrino gauge applied to (5.2) all the fields without the neutrino itself are factorized by a phase structural factor of the form $\exp(i\alpha_x)$, which expresses their behaviour upon this transform. These fields cannot be factorized by $\gamma_5$ nor can they interchange positions on (7.3) since this would lead to the vanishing or equality of masses, respectively. Therefore the general result of transforming the interaction Hamiltonian (5.2) is factorizing each term by $\exp(-i[\alpha_N+\alpha_v-\alpha_P-\alpha_e])$ and changing of factors for the dashed and un-dashed terms. Then the invariance condition is

$$C_k' = - C_k, \quad \exp(-i[\alpha_N+\alpha_v-\alpha_P-\alpha_e]) = - 1 \tag{7.7}$$

$$C_k' = + C_k, \quad \exp(-i[\alpha_N+\alpha_v-\alpha_P-\alpha_e]) = + 1 \tag{7.8}$$

Which one of the two possibilities is to choose will depend on experiment. As we shall see later the experiment gives an almost completely definite resolution to this question.

### 7.2. Two-component formalism

Dirac's equation for a zero-mass particle has the form

$$\gamma_\mu \nabla_\mu \psi_v = 0 \tag{7.9}$$

This equation is invariant with respect to the complete Lorenz group. However, it reduces to two equations for the two-component spinors $\varphi$ and $\chi$, which have the form (2.12) in the representation of Dirac-Pauli with $m_v = 0$, while in the representation

$$\alpha_{12} = \alpha_{21} = \mathbf{0}, \alpha_{11} = -\alpha_{22} = \boldsymbol{\sigma};\ \beta_{11} = \beta_{22} = 0, \beta_{12} = \beta_{21} = 1$$

they have the form

$$i\dot{\varphi} = +(\sigma \cdot p)\varphi \qquad (7.10)$$

$$i\dot{\chi} = -(\sigma \cdot p)\chi \qquad (7.11)$$

• stands for a time-derivative. These two equations are with separated variables. Each one of them is not invariant on spatial reflection and because of that it was given no physicl meaning before. The aggregate of the two equations is invariant with respect to that transform, since any of them goes into the other. Now the above limitation is lifted and a possibility arises for a description of the neutrino by means of a two component equation.

In Dirac-Pauli's representation the spinor function $\psi$ for the free states of a particle and an antiparticle can formally be arranged in a matrix of the form

$$\psi(p.m)_{11} = \psi(p.m)_{22} = 1,\ \psi(p.m)_{21} = -\psi(p.m)_{12} = \sigma \cdot p/(W+m) \qquad (7.12)$$

with (2.14-15), the normalization factor omitted. This matrix of neutrino states goes to

$$\psi_v(p)_{11} = \psi_v(p)_{22} = 1,\ \psi_v(p)_{21} = -\psi_v(p)_{12} = \sigma_p \qquad (7.13)$$

where $\sigma_p = \sigma \cdot p/p$. If we chose the z-axis in the direction of $p$, we get the following possible states:

for neutrino:                                       for antineutrino:

$(1\ 0\ 1\ 0)^T$     $(0\ 1\ 0\ -1)^T$        $(-1\ 0\ 1\ 0)^T$     $(0\ 1\ 0\ 1)^T$
spin along $p$      spin counter $p$         spin along $p$      spin counter $p$

This indicates that for a zero-mass particle two spin states are possible: along the particle momentum and counter the momentum. A similar situation occurs for the antiparticle as well. Therefore, each particle or antiparticle state in the four-component formalism is a mixture of two states: $\sigma_p = 1$ and $\sigma_p = -1$. This arrangement conserves parity.

Let us now denote by $\psi_v^{(+)}$ and $\psi_v^{(-)}$ the following functions

$$\psi_\nu^{(+)} = (1 - \gamma_5) \psi_\nu \tag{7.14}$$

$$\psi_\nu^{(-)} = (1 + \gamma_5) \psi_\nu \tag{7.15}$$

which correspond to the two possibilities (7.7) and (7.8).

It is easy to see by directly multiplying the matrices $(1 \pm \gamma_5)$ and (7.13) that the matrix $(1 + \gamma_5)$ projects the states (7.13) upon the second and the third ones, while the matrix $(1 - \gamma_5)$ does so upon the first and the fourth ones.

Choosing one of the two functions (7.14-15) corresponds to choosing the first or the second equation of the two component formalism of (7.10-11) as the equation of the two-component neutrino. Which one will be preferred will depend on the experiment. Each one of the functions $\psi_\nu^{(+)}$ and $\psi_\nu^{(-)}$ reduces the four component formalism (7.9) to a two component formalism in Dirac-Pauli's representation in as much as they have only two independent components:

$$\psi_\nu^{(+)} = 2^{-1/2} \left( (1 + \sigma_\nu)\varphi \quad (1 + \sigma_\nu)\varphi \right)^T$$

$$\psi_\nu^{(-)} = 2^{-1/2} \left( (1 - \sigma_\nu)\varphi \quad -(1 - \sigma_\nu)\varphi \right)^T \tag{7.16}$$

This corresponds to converting the system (2.12) into a system of equations for the sum and difference of $\varphi$ and $\chi$ which are already with separated variables.

So, the transition to the two component formalism in the theory of Dirac's particles with vanishing mass means considering only one of the possible two neutrino, respectively, antineutrino states. In the new theory the neutrino is in a state of a screw with a definite left-hand or right-hand cutting, while the antineutrino is in a screw with the opposite cutting, as illustrated below:

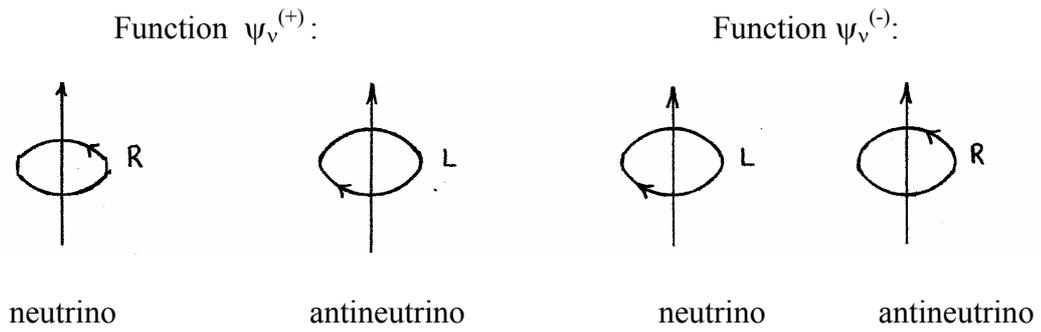

Figure 4-3:

The neutrino and antineutrino momentum and spin according to whether described by the $\psi_v^{(+)}$ function (left) or the $\psi_v^{(-)}$ function (right).

### 7.3. Invariant properties of the free neutrino field

The existence of a definite longitudinal polarization of the free neutrino leads to interesting properties of the neutrino field, as illustrated in the drawings below:

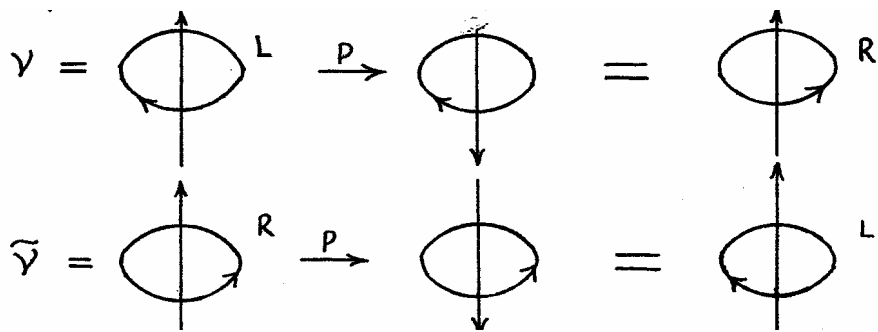

Figure 4-4:

Explaining just why neutrino-participating processes do not conserve spatial parity.

1. The neutrino does not possess any definite inherent spatial parity and in processes with its participation parity is not conserved since on spatial reflection the neutrino goes to a non-existing state (Figure 4-4).

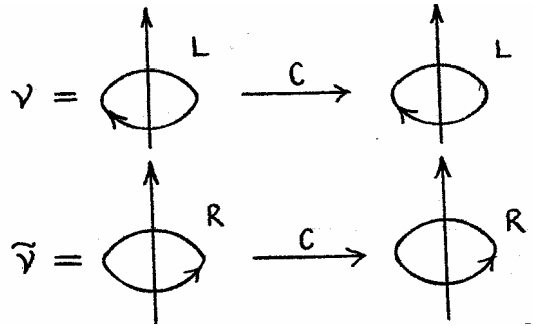

Figure 4.5:

Neutrino behaviour upon charge conjugation

2. The neutrino does not possess any definite charge parity, as well, and in processes with its participation this parity is not conserved because on charge reflection the neutrino passes into a non-existing state (Figure 4.5).

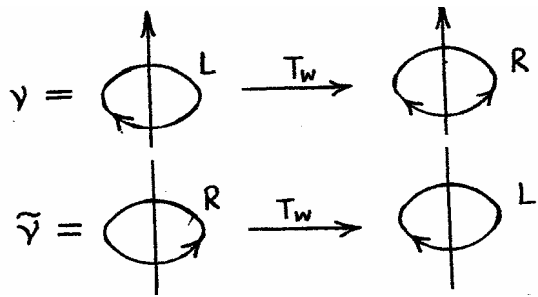

Figure 4.6:

Conceived neutrino behaviour upon time reversal

3. The neutrino. however, may possess an inherent temporal parity, in as much as on Wigner's time reversal or, equivalently, on combined inversion, it goes to an antineutrino and vice versa, as expected in case of conservation (Figure 4.6).

All that has been said about the neutrino refers to the antineutrino, as well (the second row on the illustrations).

In the two-component theory the condition $m_\nu = 0$ is necessary for the existence of a longitudinal polarization. Indeed, if $m_\nu \neq 0$ one may pass to a reference system at rest where the momentum-spin correlation for the neutrino would have lost physical meaning since $p_\nu = 0$, while on repeated passage to a moving frame would lead to a very different correlation. Thus, the requirement $m_\nu = 0$ secures the Lorenz invariance of the longitudinal neutrino state. We also remind that this state is invariant with respect to all interactions, as well, by virtue of Salam's postulate.

The functions $\psi_\nu^{(+)}$ and $\psi_\nu^{(-)}$ are eigen-functions of the operator $-\gamma_5$ :

$$(-\gamma_5)\,\psi_\nu^{(+)} = \psi_\nu^{(+)} \qquad (7.17)$$

$$(-\gamma_5)\,\psi_\nu^{(-)} = -\psi_\nu^{(-)} \qquad (7.18)$$

The eigen-values of this operator $\pm 1$ are called right-hand and left-hand chirality, respectively.

The neutrino chirality has been determined experimentally by Goldhaber, Grodzins and Sunyar [42]. They measured the circular polarization of the γ-emission following the K-capture in a nucleus originally non-polarized:

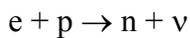

for which the only possible polarization axis following the decay is the neutrino momentum. The experiments have shown that the neutrino possesses a left-hand chirality

(left-hand screw), that is a counter momentum polarization. Then the antineutrino has a right-hand chirality (right-hand screw) - polarization along the momentum. Therefore the experiment gives preference to the condition (7.8) as regards the neutrino, as well as to the function (7.15).

## 7.4. Majorana's theory

The wave equation of the four component neutrino (mass at rest and electric charge both vanishing) (is invariant with respect to the Persy-Luders-Pauli's transform

$$\psi \to \alpha\psi + \beta\gamma_5\psi$$
$$\psi \to \delta C\underline{\psi}^T + \varepsilon\gamma_5 C\underline{\psi}^T \qquad (7.19)$$

For this reason the free neutrino state is a degenerate state, meaning that its wave function can be represented in the form

$$\psi_\nu \to a\psi + b\gamma_5\psi + cC\underline{\psi}^T + d\gamma_5\underline{\psi}^T \qquad (7.20)$$

One of the means for removing this complication is going to the two-component formalism, leading to parity non-conservation. In it, the condition for a left-hand chirality of the neutrino is a reason for describing it by means of the function:

$$\psi_\nu^{(-)} = (1 + \gamma_5)\psi_\nu \qquad (7.21)$$

while the antineutrino is being described at the same time by the charge-conjugate function:

$$C(\psi_\nu^{(-)})^T = \psi_\nu^{(+)} = (1 - \gamma_5)\psi_\nu \qquad (7.22)$$

Here the neutrino and the antineutrino are essentially different particles which agrees with the conservation law of the lepton charge.

Another way for removing the degeneracy complication has been proposed by Majorana based on the assumption that neutrino and antineutrino are identical particles. This leads to the following wave function for the neutrino:

$$\psi_\nu^M = \tfrac{1}{2}(\psi_\nu + C\underline{\psi}_\nu^T) \tag{7.23}$$

In Majorana's representation we have $C = \gamma_4^T$ and then

$$\psi_\nu^M = \tfrac{1}{2}(\psi_\nu + \psi_\nu^*) \tag{7.24}$$

This implies that Majorana's particles are described by a Hermitean (real) field.

Majorana's means conserves parity. The two theories are in a sense contradictory to each other: besides the above said, a rest mass of the neutrino field could appear in Majorama's theory as a result of virtual transitions neutrino-antineutrino, while in the two-component theory the mass is strictly vanishing at all interactions, in agreement with Salam's postulate.

## 8. Neutrino twins

### 8.1. First version of the twin theory

We have seen that setting $C_k' = +C_k$ in the general version of the β-interaction (5.6) is equivalent to the assumption for a left-hand chirality of the neutrino, while setting $C_k' = -C_k$ is equivalent to that of a right-hand chirality. At the same time, other limitations are conceivable too upon the interaction constants within the framework of the two-component formalism.

One such condition

$$C_S' = C_S,\ C_A' = C_A,\ C_V' = -C_V,\ C_T' = -C_T \tag{8.1}$$

is subject to the theory of the two-component neutrino twins.[39]

In twin theory, 2 two-component neutrino states are being considered: ones with left-hand and right-hand chirality, as well as their respective antiparticles. One of these states, the left-hand one, participates in the scalar and axial interactions by means of the function

$$\nu_1 = \nu_L = (1 + \gamma_5)\nu \qquad (8.2)$$

while the right-hand state participates in the tensor and vector interactions by means of the function

$$\nu_2 = \nu_R = (1 - \gamma_5)\nu \qquad (8.3)$$

In this way, a left-hand particle or a right-hand particle may participate in a given process depending on the character of the interaction:

$$n \xrightarrow{S(A)} p + e + \tilde{\nu_1} \qquad n \xrightarrow{T(V)} p + e + \tilde{\nu_2} \qquad (8.4)$$

$$p \xrightarrow{S(A)} n + \tilde{e} + \nu_1 \qquad p \xrightarrow{T(V)} n + \tilde{e} + \nu_2 \qquad (8.5)$$

This theory leaves unchanged the results related to the Wu effect as well as to other purely Gamow-Teller transitions. It is amazing that the term B entering in the expression for the asymmetry coefficient (5.21) is always vanishing independent of the invariance wirh respect to the time reversal. The asymmetry itself is magnified with respect to its values in the conventional two-component theory.

The twin concept annihilates automatically Fierz's interference terms of type $b_F$ and $b_{G.T.}$. This is due to the fact that a right-hand particle could never interfere with a left-hand particle. In the two-component theory, Fierz's terms occur whenever the Coulomb interaction is accounted for.

For mixed transitions with $\Delta j = 0$ the asymmetry coefficient (5.21) depends also on a typical G.T.–F. interference term of the form

$$\text{Re}\{C_T'C_S^* + C_TC_S'^* - C_A'C_V^* - C_AC_V'^* \pm$$

$$i(Ze^2/p)(C_A'C_S^* + C_AC_S'^* - C_T'C_V^* - C_TC_V'^* \} \qquad (8.6)$$

where the ± signs refer to electron (positron) decay. In the two-component theory this term is added to Wu's effect changing the asymmetry pattern. The general coefficient of (5.21) type is now proportional to

$-|A|^2 - 2\text{Re}(AV^*)$ or $|T|^2 + 2\text{Re}(TS^*)$

in case of a left-hand spiral neutrino and AV and ST versions of the theory, respectively, and to

$+|A|^2 + 2\text{Re}(AV^*)$ or $|T|^2 + 2\text{Re}(TS^*)$

for a right-hand spiral neutrino in the same versions. If the theory is invariant with respect to time reversal, then in all cases considered the effect of the interference terms (8.6) is significant.

Experiments by Wu et al.'s [45] with positron decay of $Co^{58}$, however, have shown the absence of such terms. A possible way out within the frameworks of the two-component neutrino is by introducing complex coupling constants on the account of conserving the temporal parity or by changing the form of the β-interaction, as shown by Lee [68] and Wu *et al.* [45].

Besides, interference terms of the (8.6) type are absent in twin theory and it agrees with the experiment for both versions of the interaction.

The appearance of such terms is, however, evidenced in certain experiments, as we shall see later, which are in a clear contradiction to the twin concept.

In case of a purely Fermi transitions of the O → O type, as is the decay

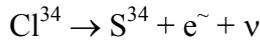

$Cl^{34} \to S^{34} + e^{\sim} + \nu$

the positron polarisation is proportional to the expression

$$\text{Re}\{C_S C_S'^* - C_V C_V'^*\} \qquad (8.7)$$

In twin theory this expression has the the form

$|C_S|^2 + |C_V|^2$

i.e. the versions S and V enter fully symmetrically. Therefore both ST and VA can lead to

correct results. In two-component theory we have two cases depending on the neutrino chirality:

$+|C_S|^2 - |C_V|^2$ for left-hand spiral neutrino

$-|C_S|^2 + |C_V|^2$ for right-hand spiral neutrino

In as much as the sign of the longitudinal polarization of the positrons is positive, the two-component neutrino agrees with the experiment at VA version and left-hand winding and at ST version with right-hand winding.

A good agreement of the twin theory with the experiment is obtained in the experiments of Telegdi et al.[38] on the β-decay of polarized neutrons. For the ratios of the intensities of the β-rays parallel and anti-parallel to the neutron spin they find the value 0.62 ±0.10. By certain corrections of the solid angle of the β–detector and the electron velocity (0.80 on the average) the distribution in reference to the electron spin has the form

$W(\theta) = 1 - (0.37 \pm 0.11)(v/c)\cos\theta$

At the same time, the prediction of the two-component theory is: − 0.08 or − 1.00 while the twin theory gives − 0.53 .

## 8.2. Second version of the twin theory

A modification of the twin theory is obtained, if assumed that the right-hand neutrino produced in the positron decay (8.5), due to the T or V interaction, is identical with the right-hand antineutrino produced in the electron decay (8.4) by S or A version, that is,

$\nu_2 \equiv \tilde{\nu_1}$ (8.8)

Now two neutrinos remain to exist: a right-hand $\nu_2 \equiv \tilde{\nu_1} \equiv \nu_R$ and left-hand $\nu_1 \equiv \tilde{\nu_2} \equiv \nu_L$. The former will be referred to as simply neutrino $\nu_R = \nu$, the latter one as antineutrino $\nu_L = \tilde{\nu}$. They are obviously charge-conjugate to each other.

In this version of the twin theory, reactions (8.4-5) take the form

$$n \xrightarrow{S(A)} p + e + \nu \qquad n \xrightarrow{T(V)} p + e + \tilde{\nu} \qquad (8.9)$$

$$p \xrightarrow{S(A)} n + \tilde{e} + \tilde{\nu} \qquad p \xrightarrow{T(V)} n + \tilde{e} + \nu \qquad (8.10)$$

Therefore, in T and V interactions a particle and an antiparticle are created, while in the S and A versions two particles or two antiparticle are created or annihilated. In this way, versions S and A violate the conservation law for the lepton number.

In agreement with this law are both the usual two-component theory and the unmodified twin theory, while Majorana's concept is directly contradictory since assuming $\nu \equiv \tilde{\nu}$.

Clearly, in certain aspects the new theory is like the two-component one, while in others it is remindful of Majorana's theory. However, it removes many of the objections usually raised to the real theory.

In conventional Majorana theory a process of the type

$$\nu + Cl^{37} \rightarrow A^{37} + \tilde{e} \qquad (8.11)$$

is completely possible. The cross-section is $5.2 \times 10^{-45}$ cm$^2$ as calculated by this theory. However, Davies [69] has obtained experimentally $0.9 \times 10^{-45}$ cm$^2$ exposing $Cl^{37}$ nuclei to a beam of particles generated in the reactor by the process

$$n \rightarrow p + e + \nu \qquad (8.12)$$

This result shows that in each case reactions of the (8.11) kind are non-observable and consequently in disagreement with the real theory.

According to the modified twin theory the reactor emits both neutrinos and antineutrinos alike. In as much as reaction (8.11) is of a mixed kind $\Delta j = 0$, the interaction has to be built up of ST or VA, e.g. of VA. The nuclei $Cl^{37}$ and $A^{37}$ are not mirror nuclei and hence the selection rules for the isotopic spin annihilate Fermi's matrix element $M_F$. Then, only the version A remains and reaction (8.11) may materialize with right-hand neutrinos coming from the reactor. Such neutrinos,, are created therein by scalar and axial β-decays. as follows:

$$n \xrightarrow{S(A)} p + e + \nu$$

If they constitute half the full flow of neutrinos, then the cross-section of reaction

$$\nu + Cl^{37} \xrightarrow{A} A^{37} + \tilde{e},$$

as calculated by modified twin theory, will be $2.6 \times 10^{-45}$ cm$^2$.

However, it is possible that there exists in nature such an asymmetry that the number of left-hand neutrinos exceeds that of the right-hand neutrinos. Using Davies's cross section, we conclude that the quantity of right-hand neutrinos amount to 0.36 of the full number of neutrinos produced in the nuclear pile. Therefore it is not surprising that the process (8.11) is non-observable.

To the contrary, Davies's experiments can be considered as proving that there exists a left-to-right asymmetry in nature.

In Majorana's theory another process is possible of the type

$$n_1 \to p_1 + e_1 + \nu \qquad \nu + n_2 \to p_2 + e_2 \qquad (8.13)$$

called a *double β-decay*. In the double β-decay two electrons are emitted simultaneously and the charge of the nucleus is changed by two units. Cowan and Reines [43] have shown experimentally that the lifetime of nuclei in the double β-decay is $10^{18}$ years, while the real theory estimates it at $10^{15}$ years. In twin theory the reactions (8.13) are conceivable subsequently in A and V versions. It is shown that it does not contradict the practical non-observability of the double β-decay.

## 9. Universal Fermi interaction

### 9.1. Universal coupling

The theory of the two-component neutrino explained in a natural way the violation of the invariance on reflection in many of the weak-interaction processes. Despite the great plausibility of the idea of a longitudinally polarized neutrino as well as its considerable success in the quantitative description of the phenomena, it is in no state to tackle such parity non-conservation processes which run without neutrino participation, and it is also far from solving the problem of which version of the β-interaction is to be chosen.

However, there exists an universal means for the description of the parity non-conservation in weak interactions which overcomes the mentioned difficulties. It has first been proposed by Sudarshan and Marshak [23] and later Feynman and Gell-Mann [21] arrived at the same conclusion by way of stringent theoretical considerations.

In their work the former two made a detailed qualitative analysis of the experimental data on the weak interactions from the viewpoint of the theory of the two-component neutrino and the conservation law of the lepton number. This analysis indicated that all phenomena related to the β-decay of nuclei and the decay of μ, π, and K mesons are well described by a mixture of vector and axial-vector versions of the Fermi interaction with the same interaction constants, that is,

$$H^i = G(\underline{\psi}_A \gamma_\mu (1+\gamma_5) \psi_B)(\underline{\psi}_C \gamma_\mu (1+\gamma_5) \psi_D) + h.c. \tag{9.1}$$

In as much as $\gamma_5$ and $\gamma_\mu$ anti-commutate, this Hamiltonian can be written in the form

$$H^i = G(\underline{\psi}_A' \gamma_\mu \psi_B')(\underline{\psi}_C' \gamma_\mu \psi_D') + h.c. \tag{9.2}$$

where $\psi_A'$, $\psi_B'$, $\psi_C'$, $\psi_D'$ are two-component fields

$$\psi_A' = 2^{-1/2} (1+\gamma_5)\psi_A = a\psi_A, \text{ etc.} \tag{9.3}$$

Therefore, the vector and axial versions connect Fermi fields with the same chirality, which is left-handed in the case of a V±A mixture. At the same time, the tensor, scalar and pseudo-scalar variants connect Fermi fields with opposite chirality two by two, since otherwise the expressions of the (9.2) type would vanish.

Thus, the universal V±A version is the only Fermi interaction which does not conserve parity though conserving chirality. This leads automatically to the two-component neutrino and a maximum asymmetry in the weak interactions. The ± sign is determined in accordance with the requirements of experiment.

### 9.2. Feynman–Gell-Mann's theory

As shown by Feynman and Gell-Mann, the universal Fermi interaction has a stringent theoretical foundation.

The spinor wave function in Dirac's equation

$$\{\nabla^{\wedge} - ieA^{\wedge} + m\}\psi(x^{\wedge}) = 0 \qquad (2.21)$$

is a four-component one. The usual explanation is that the electron has four degrees of freedom, two signs of energy and two spin states. However, this argument is not satisfactory. In Klein-Gordon's equation

$$\{\nabla_{\mu}\nabla_{\mu} - \mu^2\}\varphi(x^{\vee}) = 0 \qquad (9.4)$$

describing a particle with spin 0 the wave function has only one component rather than two. This equation is second-order and the energy sign is selected as appropriate initial conditions are imposed on $\varphi$ and $\varphi\bullet$. They predetermine the subsequent development of the particle in space-time in states with positive or negative energy.

It has proved that Dirac's equation can also be reduced to a second-order equation as the number of components of the spinor wave function is reduced by two.

For this purpose we set

$$-\psi = (1/m)\{\nabla^{\wedge} - ieA^{\wedge} - m^2\}\chi \qquad (9.5)$$

and substituting in Dirac's equation we find that the four-component wave function satisfies the following second-order equation

$$\{(\nabla_{\mu} - ieA_{\mu})^2 - \tfrac{1}{2}ie\sigma_{\mu\nu}F_{\mu\nu} - m^2\}\chi = 0 \qquad (9.6)$$

where

$$\sigma_{\mu\nu} = \tfrac{1}{2}\{\gamma_{\mu},\gamma_{\nu}\}_{-}, \quad F_{\mu\nu} = \nabla_{\mu}A_{\nu} - \nabla_{\nu}A_{\mu} \quad \text{and} \quad -\tfrac{1}{2}ie\,\sigma_{\mu\nu}F_{\mu\nu} = \boldsymbol{\sigma}\cdot(\mathbf{H} + i\mathbf{E})$$

In the theory of q-numbers it can be obtained by the Lagrangian :

$$L = (1/2m)\, \nabla_\mu \bar{\psi}\, \gamma_\mu \gamma_\nu \nabla_\nu \psi - \tfrac{1}{2} m\bar{\psi}\psi \tag{9.7}$$

by the usual gauge-invariant interchange

$$\nabla_\mu \to \nabla_\mu - ie\, A_\mu$$

to account for the electromagnetic interaction.

This Lagrangian is invariant, up to the sign (which does not change the equation of motion) with respect to the transformation of the wave function

$$\psi \to \gamma_5 \psi \tag{9.8}$$

As in the analogical case of the four-component neutrino (7.9), it now follows that equation (9.6) has two types of solutions depending on the action of the operator upon them

$$(-\gamma_5)\chi_- = -\chi_- \tag{9.9}$$

$$(-\gamma_5)\chi_+ = +\chi_+ \tag{9.10}$$

To each class of these solutions there may be set in a 1-to-1 correspondence the manifold of solutions of Dirac's equation. Indeed, if we multiply (9.5) by $(1 \pm \gamma_5)$ and make use of (9.9-10) we get

$$\chi_- = \tfrac{1}{2}(1 + \gamma_5)\psi \tag{9.11}$$

$$\chi_+ = \tfrac{1}{2}(1 - \gamma_5)\psi \tag{9.12}$$

The choice of a definite class of functions (9.9-10) corresponds to the transition to the two-component formalism in electrodynamics. It can be made only on the basis of experiment.

To find out the physical meaning of the representation (9.10) we shall use the expression (2.14) for a free Dirac spinor. If we choose more specifically function (9.9) we easily get

$(1 + \gamma_5)\psi = ( [1 - \boldsymbol{\sigma}\cdot\boldsymbol{p}/(W+m)] \quad - [1 - \boldsymbol{\sigma}\cdot\boldsymbol{p}/(W+m)] )^T \varphi$  (9.13)

Obviously a fully longitudinally-polarized electron will be described by the two-component functions ( $\boldsymbol{e} = \boldsymbol{p}/p$ )

$(1 + \boldsymbol{\sigma}\cdot\boldsymbol{e})\varphi$  (9.14)

if polarizarion is along momentum and

$(1 - \boldsymbol{\sigma}\cdot\boldsymbol{e})\varphi$  (9.15)

if polarization is counter momentum. Then we may expand the operator entering into (9.13) into two operators

$[1 - \boldsymbol{\sigma}\cdot\boldsymbol{p}/(W+m)] = a(1 + \boldsymbol{\sigma}\cdot\boldsymbol{e}) + b(1 - \boldsymbol{\sigma}\cdot\boldsymbol{e})$  (9.16)

The squares of the quantities

$a = \tfrac{1}{2}[1 - p/(W+m)]$, $b = \tfrac{1}{2}[1 + p/(W+m)]$  (9.17)

determine the part of electrons with polarization along and counter momentum, respectively. Then, the expression

$P = (a^2 - b^2) / (a^2 + b^2)$  (9.18)

will give the average polarization of the electrons in a state with eigenfunction (9.11). $P$ is easily found to be

$P = - p/W = -\beta$  (9.19)

Therefore, an electron described by the eigen-function (9.11) will have in β-decay the average polarization counter momentum of magnitude $\beta = v_e / c$. At $\beta \to 1$ the electron is fully longitudinally-polarized.

In as much as the experiment indicates that the electrons in β-decay are polarized counter momentum, we will in future use the representation (9.11), that is, in agreement with the experiment we will regard the electron as a left-handed particle.

If φ₁ and φ₂ are two-component spinors entering in ψ then the function $\chi_-$ can be written in the form

$$\chi_- = (\omega \quad -\omega)^T \qquad (9.20)$$

where ω = ½ (φ₁–φ₂). Then equation (9.6) turns into a second-order equation for thr two-component wave function ω :

$$\{(\nabla_\mu - ieA_\mu)^2 + \boldsymbol{\sigma}\cdot(\mathbf{H} + i\mathbf{E}) - m^2\}\omega = 0 \qquad (9.21)$$

In all classic electrodynamic problems where electrons are neither created nor annihilated without positrons, the theory built upon (9.21) leads to the same results as Dirac's theory. However, in describing processes like the β-decay, in which electrons alone (or positrons alone) are created, using equation (9.21) in lieu of (2.21) can essentially alter things.

Indeed, under these circumstances the general expression for the Fermi interaction (4.4) may be built up in two ways.

For that purpose, if we make use of the operator

$$-(1/m)\{\nabla^\wedge - ieA^\wedge - m\}\chi_- \qquad (9.22)$$

derivatives of ω will enter the Hamiltonian. Besides, it conserves parity at $C_k' = 0$.

It would be wise to require that the interaction would not contain gradient terms. Then it should be constructed by means of the operator

$$\tfrac{1}{2}(1 + \gamma_5)\psi = a\psi \qquad (9.23)$$

which yields

$$\Sigma_k C_k (\underline{a\psi_A}\, \Omega_k\, a\psi_B)(\underline{a\psi_C}\, \Omega_k\, a\psi_D) + h.c. \qquad (9.24)$$

Expression (9.24) leads automatically to parity non-conservation. In as much as the latter is an established experimental fact, it follows that in constructing the interaction the operator (9.23) must be regarded as more fundamental than (9.22).

Expression (9.24) can be worked out further. In so far as

$$\underline{a}\psi = \overline{\psi}\,a, \quad a = \tfrac{1}{2}(1 - \gamma_5) \tag{9.25}$$

we can write

$$\Sigma_k C_k (\overline{\psi}_A \, \underline{a} \, \Omega_k \, a\psi_B)(\overline{\psi}_C \, \underline{a} \, \Omega_k \, a\psi_D) + h.c. \tag{9.26}$$

It is not hard to see that

$$\underline{a}\,\Omega_k\,a = \Omega_k\,\underline{a}\,a = 0 \ (k = S,T,P), \quad \underline{a}\,\Omega_k\,a = \Omega_M\,\underline{a}\,a = \Omega_M\,a \ (M = V,A) \tag{9.27}$$

and then

$$\Sigma_M C_M (\overline{\psi}_A \, \Omega_M \, a\psi_B)(\overline{\psi}_C \, \Omega_M \, a\psi_D) + h.c. \tag{9.28}$$

Besides, in as much as $\gamma_5\, a = a,$ the vector and axial interactions are equivalent to each other. Then we obtain finally

$$\sqrt{8}G(\overline{\psi}_A \, \gamma_\mu \, a\psi_B)(\overline{\psi}_C \, \gamma_\mu \, a\psi_D) + h.c. \tag{9.29}$$

This interaction is to some degree unambiguous. Thus, for the process n → p + e + ṽ Hamiltonian (9.29) has the form

$$\sqrt{8}G(\overline{P}\,\gamma_\mu\,aN)(\overline{e}\,\gamma_\mu\,a\nu) + h.c. \tag{9.30}$$

while for the reaction p + ν → n + e, it assumes the form

$$\sqrt{8}G(\overline{P}\,\gamma_\mu\,\underline{a}N)(\overline{e}\,\gamma_\mu\,a\nu) + h.c. \tag{9.31}$$

According to the general considerations presented in Section 4, these two expressions describe the same process. The ambiguity mentioned can be removed by comparing the results made based on (9.30) and (9.31) with the experimental results. On decay of polarized neutrons, Hamiltonian (9.30) leads to a minimal asymmetry in the angular distribution of the electrons, while (9.31) leads to maximum asymmetry. In the theory of the two-component neutrino the two Hamiltonians correspond respectively to the V+A and V−A versions of the interactions.

So, the theory by Feynman and Gell-Mann has led to the same interaction Hamiltonian as have Sudarshan and Marshak in quite a different way. This law is in harmony with the experiment not only qualitatively but also quantitatively. For instance, the μ-meson lifetime computed by the theory amounts to

$$\tau = (192\pi^3/G^2\mu^5) = (2.26 \pm 0.04) \, \mu s$$

provided the coupling constant is $G = (1.01 \pm 0.01) \times 10^{-5} \, M^{-2}$ where M is the proton mass at $h = c = 1$. The experimental value agrees within 2% with the above one. The theory also explains the parity non-conservation in $2\pi$ or $3\pi$ decay of the K meson which proceed with no participations of neutrino – perhaps this is one of its best virtues. Lately, it gained new confirmations from experiment, the most significant one being the experimental observation of the electronic decay of the π-meson.

## 10. Comparison with experiments

We have seen that as a result of the non-conservation of parity the γ–rays emanated from the products of the β–decay of oriented nuclei will be circularly polarized, where the degree of polarization will be proportional to the cosine of the angle between the momentums of the γ–quantum and the β–particle. The circular polarization of the γ–quanta can be analyzed by means of a cylindrical electromagnet magnetized to saturate parallel or anti-parallel to their propagation. The principle of this analysis is based on the existence of a spin-dependent part of the Compton scattering of circularly polarized photons.

The correlation beta-polarization-gamma considered by us above for the most frequently occurring decay of the form

$$j \xrightarrow{\text{(allowed } \beta\text{-transition)}} j' \xrightarrow{\text{(2L-pole } \gamma\text{-ray)}} j'' = j' - L$$

can be expressed in the form ($\tau = \pm 1$)

$$W(\theta,\tau) = 1 - \tau A(v/c)\cos\theta \qquad (10.1)$$

where at j=j'=j" A is easily obtained by (6.13) for $\varphi = 0$, while in the general case it has a non-essentially altered form.

For the first time Shopper has used this method to verify the parity non conservation in the electronic decay of $Co^{60}$ and positron decay of $Na^{22}$. He obtained the following values for the asymmetry coefficient A at these two purely Gamow-Teller transitions:

| | | | |
|---|---|---|---|
| $Co^{60}$ | $\Delta j = -1$ | $A = -0.41 \pm 0.07$ | theoretical $-0.33$ |
| $Na^{22}$ | $\Delta j = +1$ | $A = +0.39 \pm 0.08$ | theoretical $+0.33$ |

Aside are given for comparison the theoretical calculations carried out using the theory of the two-component neutrino. Both series agree beautifully for left-handed neutrino in axial version, as well as for right-handed neutrino in tensor version. We should, however, stress that the accuracy of the experiment does not exceed 20%. By the same convention they are in accord with the twin theory.

Similar results have emerged in the experiments of Boem & Vapstra and those by Debruner & Cundig. The latter two authors have obtained an even better agreement with the theory within experimental error:

| | | | |
|---|---|---|---|
| $Co^{60}$ | $\Delta j = -1$ | $A = -0.344 \pm 0.09$ | theoretical $-0.33$ |

In another series of experiments Boem and Vapstra have investigated a great number of compounds:

| | | | |
|---|---|---|---|
| $Co^{60}$ | $\Delta j = -1$ | $A = -0.41 \pm 0.08$ | theoretical $-0.33$ |
| $Na^{24}$ | $\Delta j = 0$ | $A = +0.07 \pm 0.04$ | theoretical $+0.08$ |
| $Sc^{44}$ | $\Delta j = 0$ | $A = -0.02 \pm 0.04$ | theoretical $-0.17$ |
| $Sc^{46}$ | $\Delta j = 0$ | $A = +0.33 \pm 0.04$ | theoretical $+0.08$ |
| $V^{48}$ | $\Delta j = 0$ | $A = +0.06 \pm 0.05$ | theoretical $+0.08$ |
| $Co^{58}$ | $\Delta j = 0$ | $A = -0.14 \pm 0.07$ | theoretical $-0.17$ |

The theoretical calculations are made on the assumption for a Gamow-Teller character of the transitions and a two-component neutrino – right-handed in tensor version and left-handed in axial version. A clear-cut distinction from experiment is obtained in cases like $Sc^{44}$ and $Sc^{46}$. This indicates strong interference effects of the GT-F type.

The appearance of similar interference terms has also been observed in an earlier experiment by Boem and Vapstra with $Sc^{46}$ and in those of Apel and Shopper with $Sb^{124}$:

$Sb^{124}$ $\Delta j = 1$ $A = -0.13 \pm 0.06$ theoretical $-0.33$

The same paper presents the results for two purely Gamow-Teller transitions:

$Co^{60}$ $\Delta j = -1$ $A = -0.35 \pm 0.05$ theoretical $-0.33$
$Zr^{95}$ $\Delta j = +1$ $A = -0.46 \pm 0.06$ theoretical $-0.33$

which agree well with the theoretical predictions. The last of them in the worst possible case speaks of a weak interference. Of a similar character is the result by Apel, Shopper and Bloom :

$Ca^{22}$ $\Delta j = +1$ $A = +0.295 \pm 0.054$ theoretical $-0.33$

and that of Stephen

$Sc^{46}$ $\Delta j = 0$ $A = -0.24 \pm 0.02$ theoretical $-0.08$

A strong Gamow-Teller – Fermi interference is observed in the decay of $Mn^{52}$ by Boem:

$Mn^{52}$ $\Delta j = 0$ $A = -0.16 \pm 0.05$ theoretical $-0.08$

The interference effects are well explained by the two-component neutrino theory in its two versions – ST and AV. They are, however, in a strong contradiction with the two-component twin theory by virtue of their very existence.

All experiments mentioned lead to conclude that for constructing a Hamiltonian for the β–interaction one has to select either a ST version with a right-hand neutrino or a VA version with a left-hand neutrino. This also follows from the fact that the S and T versions connect Fermi particles with opposite chirality, while V and A do so for Fermi particles with the same chirality, since it has been established experimentally that the β−particles possess a left-hand chirality. On the other hand, the magnitude of the asymmetry A tells in favor of excluding the combinations VT and SA. The above results can be expressed schematically in the form of a correlation between electrons and antineutrinos or between positrons and neutrinos alike in the respective β−decays.

For obtaining the neutrino chirality and thereby the version of the β−interaction, Goldhaber, Grodzins and Sunyar have used the circular polarization of the γ- quanta following K-capture

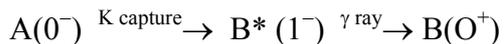

A($0^−$) $\xrightarrow{K\ capture}$ B* ($1^−$) $\xrightarrow{\gamma\ ray}$ B($O^+$)

Applying the conservation laws of momentum and angular momentum it can be verified easily that the neutrino departing forwards relative to the nuclear spin has the same chirality with the departing backwards γ- quantum. That is why the problem of determining the neutrino chirality turns into a problem for measuring the circular polarization of the γ- rays. The most appropriate nucleus for that purpose is $Eu^{152}$. The experiments conducted show that the γ- rays have a negative chirality.

The theory of the left-hand screw neutrino is for the time being in full agreement with the overwhelming part of the experiments on β−decay. This refers not only to the polarization effects. On determining the cross-section of reverse β−decay of the proton

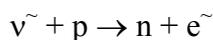

$\tilde{v} + p \rightarrow n + \tilde{e}$

this theory leads to the value $12\times10^{-44}$ $cm^2$, which responds beautifully to the requirements of experiment: $11\pm4\times10^{-44}$ $cm^2$. Besides, perhaps most important is the connection with the universal Fermi interaction which has imposed itself lately because of its great practical plausibility and considerable success.

On the other hand, the scientific thinking has been more and more approaching the conservation law for the lepton number. In favor of this law tell the practical non-observability of the double β−decay and the results of Davies' experiments even though the latter may also allow for another explanation.

In this way, it may now be safe to conclude that the theory of the two-component neutrino and the conservation law for the lepton number describe and reflect correctly the processes in nature. Yet, the existence of novel effects cannot be ruled out that have remained concealed within the interval of experimental errors. For instance, it is possible that the negligible number of $A^{37}$ nuclei (0.3 ± 3.4 per day) which has been detected by Davies in his thousand gallons of $CCl_4$ could be due to asymmetry like the one in section 8 rather than to cosmic rays. If so, why does nature prefer one kind of particles to the other? We can say following Wu that what is needed is a better experimental accuracy.